# STUDY OF NON-ALTERED COLORECTAL TISSUE BY FT-RAMAN SPECTROSCOPY

P. O. Andrade[1], R. A. Bitar[1], K. Yassoyama[1], H. Martinho[1*], A. M. E. Santo[1], P. M. Bruno[2], A. A. Martin[1]

[1]Laboratory of Biomedical Vibrational Spectroscopy, Institute of Research and Development - IP&D, University of the Valley of Paraiba - UniVap, Av. Shishima Hifumi, 2911, CEP 12244-000, São José dos Campos, SP, Brazil.

[2]Vivalle Hospital, Av. Lineu de Moura 995, São José dos Campos, SP, Brazil.

## Abstract

**Abstract** FT-Raman spectroscopy was employed to study (in-vitro) non-altered human colorectal tissues aiming evaluate their spectral differences. The samples were collected from 39 patients, adding 144 spectra. The results enable one estabilish 3 well defined spectroscopic groups of what was consistently checked by statistical (clustering) and biological (histopathology) analysis. The similarity within each group was better than 90%. Groups **1** and **2** had 80% of similarity while presenting both 70% of similarity to group **3**. All groups presented connective and epithelial tissues as histopathological characteristic. Group **1** differs from others by the presence of soft muscle while group **3** by presence of fatty tissue. Tissues from group **2** are composed by samples with connective and epithelial tissues. This study is very relevant in order to establish the non-altered or normal spectroscopic standard for future studies and applications on optical biopsy and diagnosis of colorectal cancer.

## Introduction

Cancerous growths in the colon, rectum and appendix are called colorectal cancer, colon cancer or bowel cancer. The colon and the rectum are part of the large intestine, which is part of the digestive system. Colorectal tumors may form in the lining of the large intestine. Many colorectal cancers are thought to arise from adenomatous polyps in the colon. These mushroom-like growths are usually benign, but some may develop into cancer over time. Studies show that the main factors that could increase a chance of one to develop colorectal cancer are age, polyps presence, fat-rich diet, personal history, family history and ulcerative colitis [1, 2].

The colorectal cancer is the third most common form of cancer and the second leading cause of death among cancers in the Western world [2]. Currently, in Brazil, the colorectal cancer is the fifth most common cause of death for cancer, according to National Institute for Cancer (INCA) [3]. Actually, as all kinds of cancers, the early detection of colorectal cancer is fundamental for the fast and successful treatment of the disease. The following procedures are conventionally used for its early detection: fecal occult blood test or FOBT; sigmoidoscopy; colonoscopy; double contrast barium enema or DCBE; digital rectal exam or DRE (rectal touch) and surgical biopsy for histopathological analysis [4].

The conventional biopsy of the suspicious lesion for analysis is considered the gold standard procedure for the diagnosis confirmation and classification of the illness. However, this practice presents some disadvantages. This procedure may involve some risks to the patient, mainly when the central nervous and/or the cardiovascular systems are involved. Also, inappropriate storage may lead to biochemical and structural changes of the tissue. Long-term histopathological analysis, which is a subjective method based on the pathologist experience, eventually delays early diagnostics.

Optical biopsy has been extensively studied as a proposal for the cancer diagnosis, since the light can be delivered and collected in near real time via optical fibers, allowing possible therapeutic treatment precociously. Amongst the optical techniques, the Raman spectroscopy can provide a wealth of spectrally narrow features that are related to the structural and biochemical composition of the sample under study. The analysis is based on the inelastic scattering of light by molecules that constitute the biological tissue [5, 6].

Raman spectroscopy is a non-destructive technique and does not require previous sample preparation, allowing data collection repetition or complementation. Recently, Raman Spectroscopy has been extensively used in biomedicine as a tool for the diagnosis of tissue lesions, analysis of blood components, study of single cells and others [7-11]. The advances in optical-electronic devices, mainly in detectors and lasers technology, have increased the number of applications and active research groups in this field.

Histopathologists are able to provide acceptable classification into large subgroups of epithelial cancer and non-cancer. However, while pathologists can demonstrate acceptable levels of agreement for the major comparative groups of cancer against negative, the division into subgroups of normal, mild pre-cancerous change, severe pre-cancerous change and cancer has revealed that are poorer level of agreement [12]. In this sense, Raman technique would be able to perform quantitative and qualitative analysis since it shows high sensitivity to small structural changes in biological tissues [13].

In the present work, FT-Raman spectroscopy is employed to study (*in-vitro*) non-altered human colorectal tissues. The aim is to evaluate the spectral differences of the complex non-altered or normal colon.  This work is a preliminary study to set a reference spectra database for non-altered colon tissues and experimental parameters for later comparison with displastic colorectal tissues. This method presents as a potential analytical tool for minimally invasive *in-vivo* early diagnosis of large intestine cancer.

## Methodology

This study was performed following the ethical principles established by the Brazilian Federal Healthy Ministry, according to Resolution 196/96 of the National Health Council, being approved by the Ethical Commission in Research at Univap. All the patients were informed about the research and they signed an agreement term for scientific use of the collected material.

**A. Tissues**

A total of 39 samples of colon-rectal human were supplied for this study. The samples were obtained from male and female patients with different ethnical and age characteristics. The patients were submitted to colonoscopy or resection surgery. The non-altered colorectal fragments were sampled at least 10 cm away from the free margin of the lesion site. Each tissue was sectioned into 3, 4 or 5 partitions depending on their size. The Raman spectra were measured on 3 different points on each sample adding 144 spectra. Soon after, the samples were fixed in 10 % formaldehyde solution, to further histopathological analysis.

**B. FT-Raman Spectroscopy**

A FT-Raman spectrometer (Bruker RFS 100/S) was used with an Nd:YAG laser at 1064 nm as excitation light source. The laser power at sample was kept 230 mW while the spectrometer resolution was set to 4 $cm^{-1}$. The spectra were recorded with 300 scans.

For FT-Raman data collection, samples were brought to room temperature and kept moistened in 0.9 % physiological solution to preserve their structural characteristics, and placed in a windowless aluminum holder for the Raman spectra collection. We notice that the chemical species present in the physiological solution ($Ca^{2+}$, $Na^{+}$, $K^{+}$, $Cl^{-}$, water) do not have measurable Raman signal and their presence do not affect the spectral signal of the tissues.

**C. Data Analysis**

All spectra were baseline corrected and normalized to the 1445 $cm^{-1}$ band intensity. This band corresponds to C—H deformation mode of methylene group and it is nearly conformational insensitive. For this reason, it is a good standard for biological Raman spectral normalization. The spectra were classified in groups using multivariate statistical packages of the Minitab 14.20 (Minitab Inc.) software. The results were displayed in a tree diagram (or dendogram) by their similarity level. The spectral distance was calculated by the Pearson correlation factor while the distance between two clusters was measured by the complete linkage method.

## Results and Discussion

Figure 1 shows ten randomly chosen Raman spectra of non-altered colon tissues. These spectra illustrate the intrinsic variability of the colon tissues that could induce wrong optical biopsy diagnosis. This fact reinforces the need for establishing a spectral standard for this kind of tissues. Table 1 displays the peak position and mode assignment of the main Raman bands.

Figure 2 shows the dendogram obtained for the 144 Raman spectra grouped by their similarity. It is clear the amalgamation of three distinct groups. The spectra within each formed group had similarity level better than 90 %. Groups **1** and **2** had 80% of similarity while presenting both 70% of similarity to group **3**.

To establish a characteristic Raman spectra of each group were taken the average of those spectra presenting similarity level greater than 95% within group. The results are presented on Figure 3. It is clear the great spectral resemblance among groups **1** and **2**. The bands of aminoacids, proteins, collagen, and lipids have similar intensities in these groups while the group **3** presents an overall spectral intensity decrease.

In order to verify whether these spectral grouping had some biological meaning the histological and spectral data were confronted. The histological section of each tissue piece were the Raman spectroscopy were taken was analyzed looking for similar characteristics. Group **1** showed the presence of epithelial layer, connective tissue papillae, and smooth muscle tissue as can be observed in Figure 4a. Group **2** is composed of tissues with epithelial layer and connective tissue papillae, as can be observed in Figure 4b. It was observed that group **3** presented mostly fatty and slack conjunctive tissue from 70 years old patients. A representative histological section for this group is shown in the Figure 4c.

Both, from spectroscopy and histopathologly point of view, the groups **1** and **2** showed very similar characteristics. The difference between these groups was the presence of the smooth muscle tissue. The group 1 samples are muscle-rich because they were sampled by deeper surgical procedure. The overall spectral decreasing seen in the characteristic spectra (Fig. 3c) of group **3** could be originate by the tissue degenerative process pointed out on this group by the histological analysis.

## Conclusion

The results presented in this work indicate that one could estabilish 3 well defined spectroscopic groups of non-altered colorectal tissues. These grouping was consistently checked by statistical (clustering) and biological (histopathology) analysis. The similarity within each group was better than 90%. Groups **1** and **2** had 80% of similarity while presenting both 70% of similarity to group **3**. All groups presented connective and epithelial tissues as histopathological characteristic. Group **1** differs from others by the presence of muscle while group **3** by presence of fatty tissue.

These results are very relevant because it reveals the existence of an understandable intrinsic spectral variability that could furnish a basis for a non-altered colorectal spectroscopy standard. The recognizing of this standard is the first step for future studies and applications of Raman spectroscopy on optical biopsy and diagnosis of colorectal cancer.

## Acknowledgments

Authors would like to thank the Colonoscopy Service of the Portuguese Charity Hospital of São José of Rio Preto, São Paulo, the Coloproctology Department of the Medicine Faculty of São José do Rio Preto (FUNFARME), where surgeries were accomplished, and the Vivalle Hospital of São José dos  the Campos, SP, for the samples supplement used in this study. This work is supported by CNPq (project 401018/2005-9) and FAPESP (project 01/14384-8).

## Figure Caption

**Figure 1**: Ten randomly chosen spectra of large intestine non-altered tissue.

**Figure 2**: Dendogram showing the classification of the large intestine non-altered tissue Raman spectra in 3 groups.

**Figure 3**: Averaged Raman spectra of groups 1 (a)), 2 (b)) and 3 (c)).

**Figure 4**: Representative histological sections of a) group **1** b) group **2** c) group **3**.

**Table 1**: Peak position and vibrational mode assignement.

| Peaks position ($cm^{-1}$) | Vibrational Mode | Major Assignments |
|---|---|---|
| ~ 550 | S—S | disulphide bridges in cysteine |
| 855-870 | ν(C—C), ring breathing, ν(O—P—O) | proline, tyrosine, DNA |
| 900-980 | ν(C—C), α-helix | proline, valine, protein conformation, glycogen |
| ~ 1000 | symmetric ring breathing mode | phenylalaline |
| 1080-1100 | ν(C—C) or ν(C—O), ν(C—C) or ν($PO_2$), ν(C—N), ν(O—P—O) | lipids, nucleic acids, proteins, carbohydrates |
| 1260-1280 | ν(C—N) of amide III, ν(=C—H) | proteins (α-helix), lipids |
| 1300-1310 | δ($CH_2$), δ($CH_3CH_2$) | adenine, cytosine, collagen, lipids |
| ~ 1450 | δ($CH_2$) | lipids, carbohydrates, proteins and pentose |
| 1650-1660 | ν(C=O) of amide I, ν(C=C) | proteins (α-helix), lipids |
| ~ 1750 | ν(C=O) | lipids |

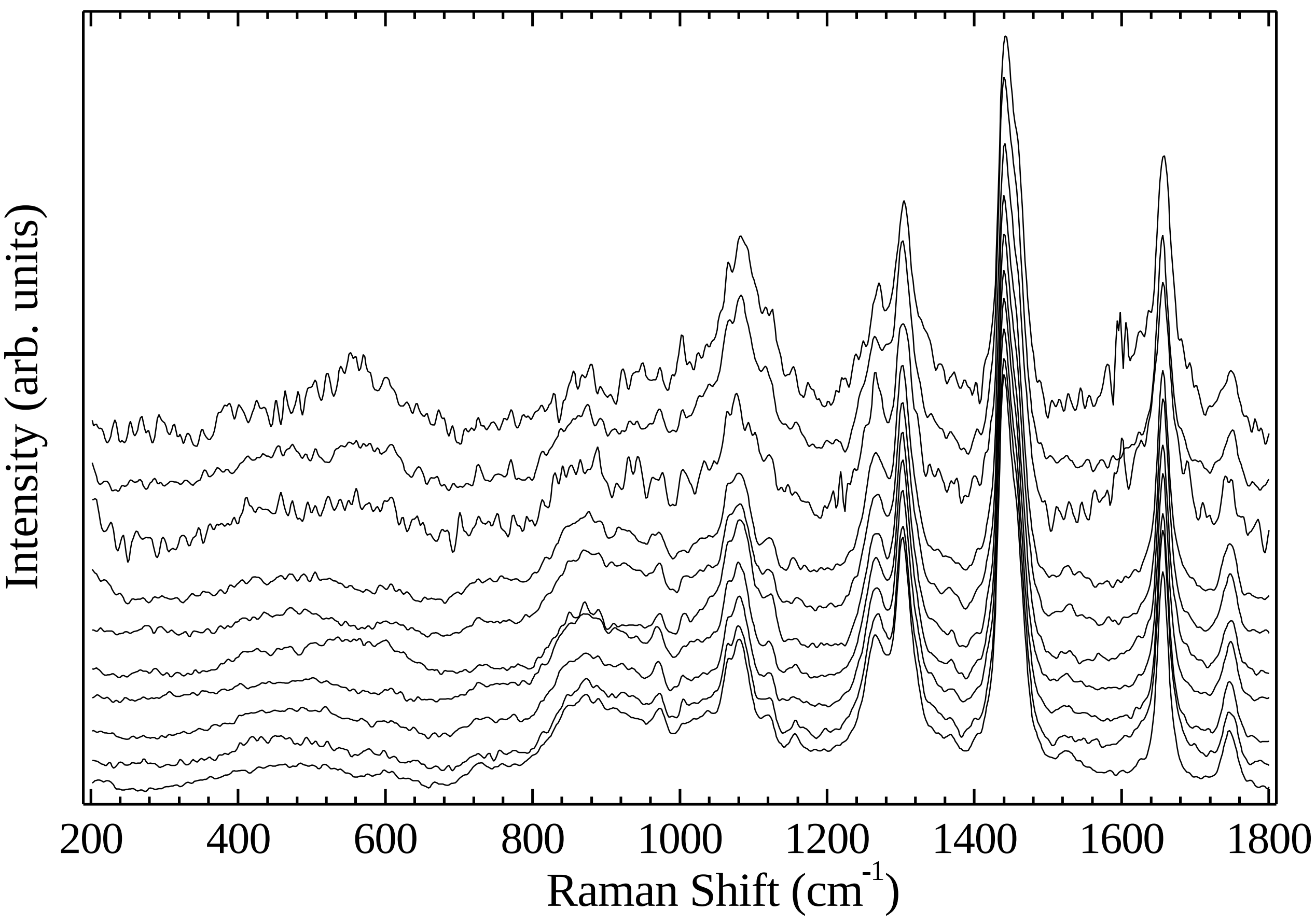

**Figure 1**

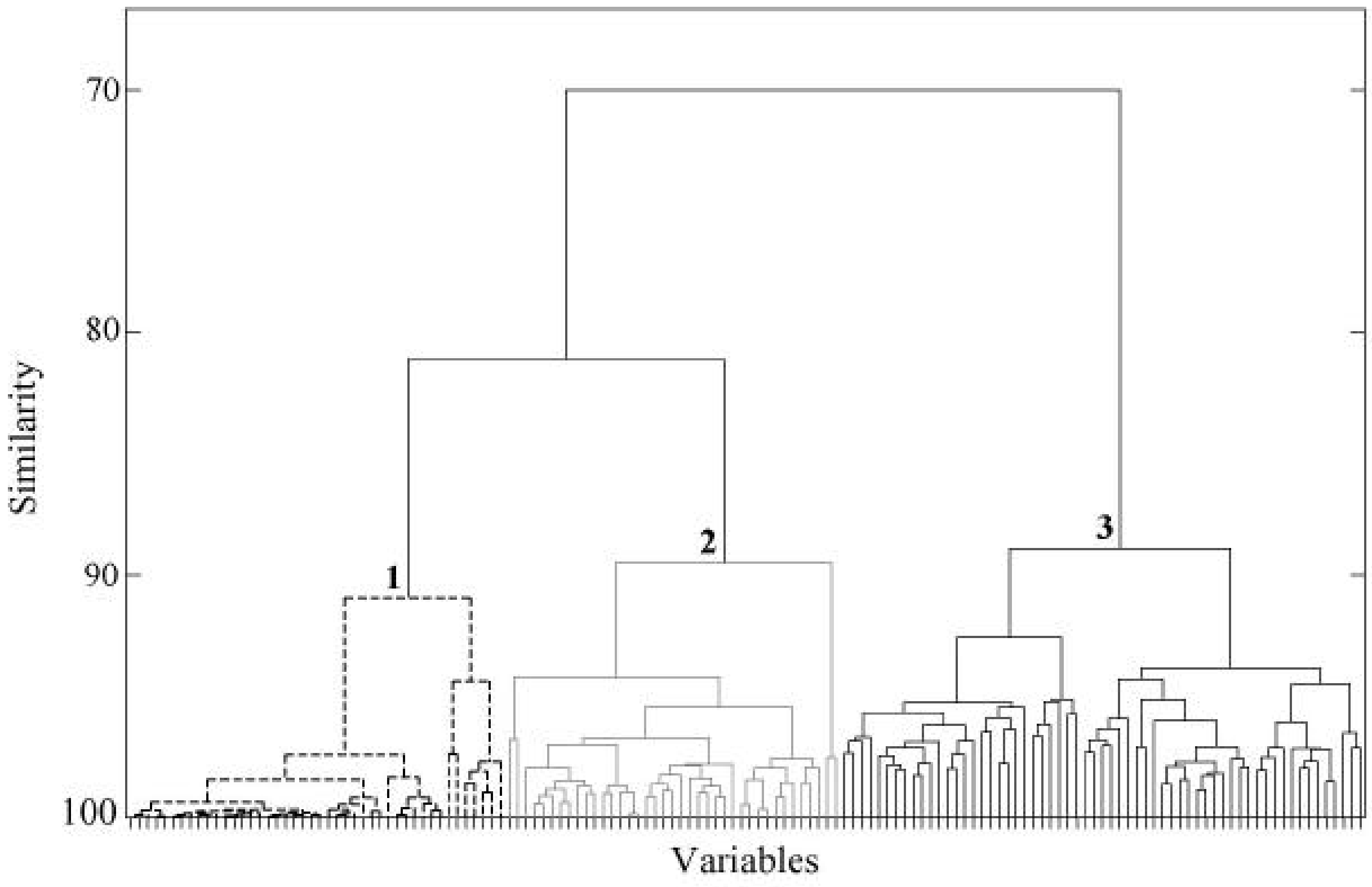

**Figure 2**

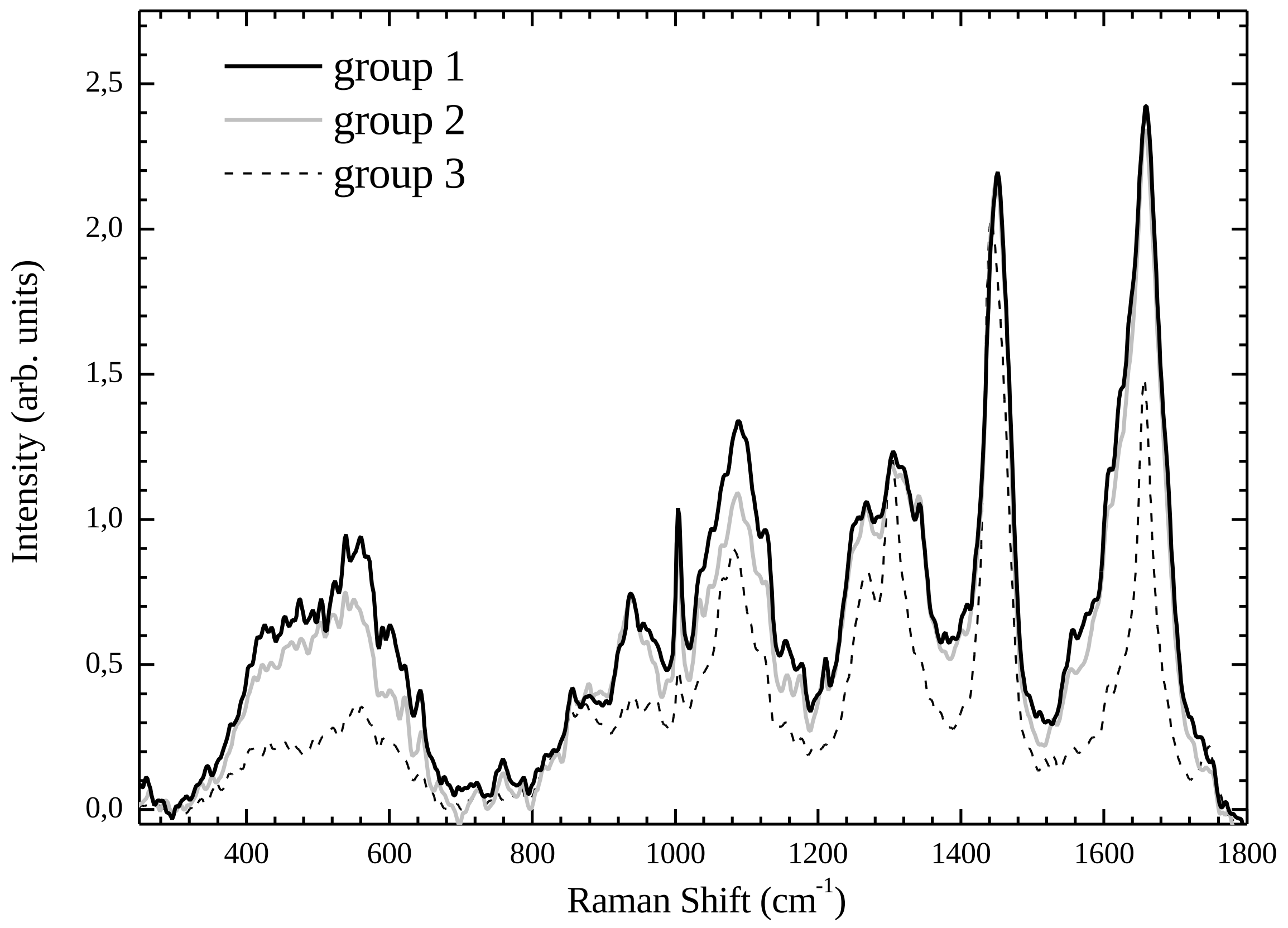

**Figure 3**

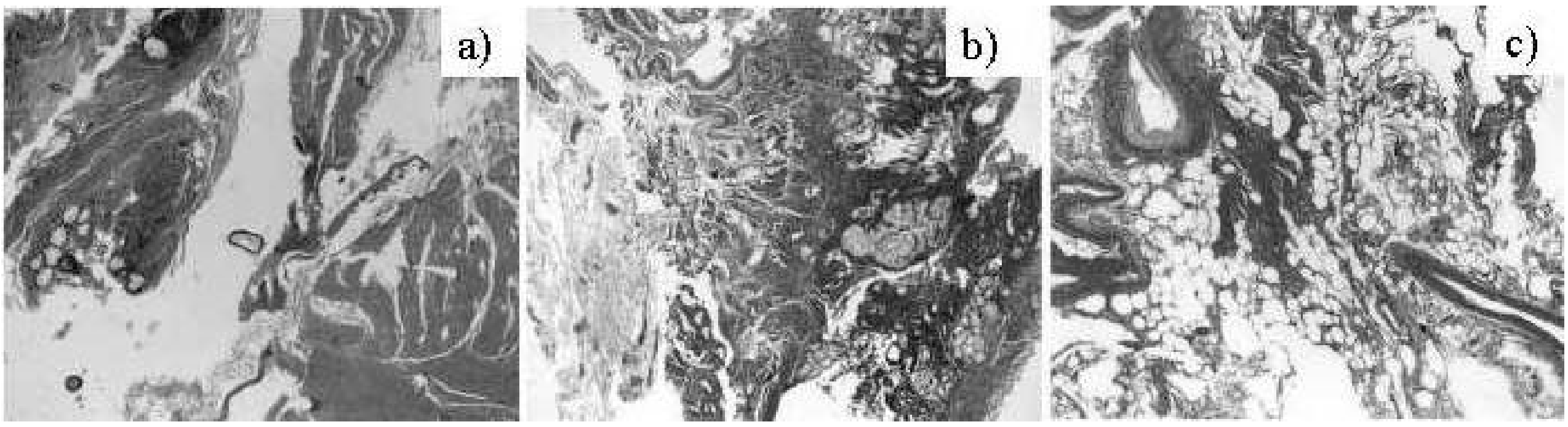

**Figure 4**